\documentclass[preprint,showpacs,aps,draft]{revtex4}

\usepackage{graphicx}
\usepackage{dcolumn}
\usepackage{bm}

\def\Journal#1#2#3#4{{#1} {\bf#2}, {#3} {(#4)}}

\def\NP{{ Nucl. Phys.} }
\def\PLB{{ Phys. Lett.}  B}

\def\PPNP{{ Prog. Part. Nucl. Phys.}}
\def\PRP{{ Phys. Rep.}}
\def\PRL{ Phys. Rev. Lett.}

\def\PRD{{ Phys. Rev.} D}
\def\PRC{{ Phys. Rev.} C}

\def\EPJC{{Eur. Phys. J.} C}

\def\MPLA{{Mod. Phys. Lett.} A}
\def\CPC{Comput. Phys. Commun.}

\def\ra{\rightarrow}

\def\be{\begin{equation}}
\def\ee{\end{equation}}
\def\bea{\begin{eqnarray}}
\def\eea{\end{eqnarray}}

\def\ua{\uparrow}
\def\da{\downarrow}
\def\qbar{{\bar q}}
\def\ubar{{\bar u}}
\def\dbar{{\bar d}}
\def\sbar{{\bar s}}

\def\NP{{ Nucl. Phys.}}
\def\ANP{{Adv. Nucl. Phys.}}

\begin{document}
\title{
Non-perturbative structure of the polarized nucleon sea}
\author{Fu-Guang Cao}
\email{f.g.cao@massey.ac.nz}
\author{A. I. Signal}
\email{a.i.signal@massey.ac.nz}
\affiliation{Institute of Fundamental Sciences PN461 \\ Massey University \\
Private Bag 11 222,  Palmerston North \\
New Zealand}

\vskip 0.5cm
\begin{abstract}
We investigate the flavour and quark-antiquark structure of the polarized 
nucleon by calculating the parton distribution functions of the nucleon sea
using the meson cloud model. 
We find that the SU(2) flavor symmetry in the light antiquark sea and quark-antiquark symmetry
in the strange quark sea are broken, {\it i.e.} $\Delta\ubar < \Delta \dbar$ and
$\Delta s < \Delta \sbar$.
The polarization of the strange sea is found to be positive, which is in contradiction to
previous analyses.
We predict a much larger quark-antiquark asymmetry in the polarized strange
quark sea than that in the unpolarized strange quark sea.
Our results for both polarized light quark sea and polarized strange quark sea
are consistent with the recent HERMES data.

\end{abstract}



\maketitle

\setcounter{footnote}{0}
\section{Introduction}


Since the famous EMC experiment \cite{EMC} at CERN in 1988 there has been 
great interest in the polarized quark distributions of the proton and neutron.
The reason for this interest in the spin dependent structure of 
the nucleon is that the EMC experiment (and other polarized DIS 
experiments \cite{others}) could be interpreted as showing that the quarks 
carry only a small proportion of the total angular momentum of the 
proton. 
A further conclusion, made using SU(3) flavour arguments, is that 
the strange sea quarks in the proton are strongly polarized opposite 
to the polarization of the proton \cite{JEllisMK}.
Both these results are much at odds with expectations based on 
constituent quark models of the nucleon. 
Further experiments have generally confirmed the EMC results for the 
proton - photon asymmetry $A_{1}$ and proton spin structure function 
$g_{1}(x)$, as well as extending these to deuteron and
neutron targets \cite{A1g1_SMC}
and to the second nucleon - photon asymmetries $A^{N}_{2}$ and the 
related structure functions $g^{N}_{2}(x)$ \cite{A2g2_E155}. 

All these experiments have been inclusive scattering using the virtual 
photon as the probe of nucleon structure. 
Unfortunately no experiments have been performed using neutrino 
scattering, where one of the vector bosons is the probe. 
Thus there is no information on the spin dependent analogue of the 
unpolarized structure function $F_{3}(x)$. 
This makes the decomposition of the measured structure functions into 
`valence' and `sea' parts very difficult, and flavour decomposition 
much harder to do than in the unpolarized case. 
However the HERMES collaboration \cite{HERMES02} has recently reported the first 
results from its semi-inclusive scattering measurements, where final 
state pions and kaons are measured.
This provides a method to extract the spin dependent quark 
distribution of a given flavour, supposing that we have enough information 
about the relevant fragmentation functions $D^{h}_{q}(Q^{2}, z)$. 
The HERMES collaboration's first semi-inclusive results have been 
able to extract the ratios of spin dependent to spin independent quark 
distributions 
$( \frac{\Delta u}{u}, \frac{\Delta \bar{u}}{\bar{u}}, 
\frac{\Delta d}{d},  \frac{\Delta \bar{d}}{\bar{d}}, 
\frac{\Delta s}{s}, \frac{\Delta \bar{s}}{\bar{s}}=0 \pm \frac{1}{3})$. 
In contrast to earlier flavour decompositions \cite{HERMES99,SMC98},
which have needed SU(3) flavour symmetry assumptions for the sea distributions
$\Delta \ubar = \Delta \dbar = \Delta s = \Delta \sbar$,
the only symmetry assumption made in the HERMES data analysis is that
$\frac{\Delta \bar{s}(x)}{\bar{s}(x)}=0$.
In the context of the earlier results from inclusive DIS, the results for 
the spin dependent sea quark distributions $\Delta \bar{u}(x), 
\Delta \bar{d}(x), \Delta s(x)$ are rather surprising: the polarization 
of each flavour is very small, and compatible with zero. 
This may imply a breaking of SU(3) flavour symmetry.

With this new data in mind, it is important that other sources of information 
on the spin dependent sea quark distributions be examined. 
In particular model calculations of the parton distributions can provide 
some insight into whether the antiquark polarizations can be expected 
to be large or small, and whether the sign of the polarization is 
positive or negative. 
In this paper we will investigate the spin dependent sea quark 
distribution functions ($\Delta \ubar,\Delta \dbar,\Delta s, \Delta \sbar$)
in the context of the meson cloud model (MCM). 
In this model the physical nucleon wavefunction contains virtual 
meson - baryon components which `dress' the bare nucleon.
The MCM provides a natural explanation for symmetry breaking 
among the parton distributions, in particular the flavour asymmetry in 
the unpolarized nucleon sea $(\bar{d}(x) > \bar{u}(x))$ \cite{Explanations}
seen in the NMC \cite{NMC} and 
E866 \cite{E866} experiments.
Thus it is reasonable to ask whether the violation of the Ellis-Jaffe 
sum rule \cite{EJSR} can also be explained in this model. 
In previous calculations \cite{CBorosT} it has been shown that
including effects of the meson cloud significantly lowers the value of 
$\Delta \Sigma = \sum_{f} (\Delta Q_{f} + \Delta \bar{Q}_{f})$, which is the 
total spin carried by the quarks and anti-quarks. 
This arises firstly because of `dilution' of the bare proton quark distributions 
by those of the cloud and secondly from the inclusion of non s-wave components 
in the cloud wavefunction, which effectively increases the proportion of the 
proton spin due to orbital angular momentum.  
Boros and Thomas \cite{CBorosT} studied the effects of strange mesons and baryons
in the cloud by considering $\Lambda K$ and $\Sigma K$ components in
the Fock wavefunction of the proton and found that the polarization of the strange
sea was small $(\Delta S + \Delta \bar{S} < 0.01)$.
However higher mass components in the Fock wavefunction,
in particular $\Lambda K^{*}$ and $\Sigma K^{*}$, can have important effects
on the strange see.
Our recent investigation of the unpolarized strange sea \cite{FCaoS_KKstar} has shown that
although these higher mass components may be kinematically suppressed,
they have large couplings to the 
nucleon, and numerically give amplitudes of similar size to the lowest mass 
states. 
Also the $K^{*}$ is a vector meson, so any polarization of the anti-strange quark 
in the $K^{*}$ will give a contribution to the $\Delta \bar{s}(x)$
distribution of the proton, whereas there can be no such contribution from 
the $\bar{s}$ of the Kaon.

In a previous paper \cite{FCaoS_plzdsea} we have calculated the flavour asymmetry 
$\Delta \bar{u}(x) - \Delta \bar{d}(x)$ in the MCM. 
These results are consistent with the HERMES data.
Here we shall extend that calculation to consider each of the light antiquark 
flavours separately. Thus the fluctuations that contribute to $\Delta \ubar(x)$ and
$\Delta \dbar(x)$ equally and make no contributions to
$\Delta \bar{u}(x) - \Delta \bar{d}(x)$ are now included.

In addition to the MCM contribution to the light polarised antiquark 
distributions there may be other contributions to these distributions. 
These would be mainly non-perturbative, arising from the distributions in the 
bare nucleon, and would need to be calculated in some model of the bare 
nucleon {\em e.g.} the MIT bag model. 
These contributions are not necessarily small, however we shall see that the 
HERMES data does constrain the size of these bare distributions. 
Perturbative contributions to the light polarised antiquark distributions, 
which come from QCD evolution from the model scale ($Q_{0}^{2} < 1$ GeV$^{2}$) 
up to experimental scales, are expected to be small as the first moment of 
these distributions changes rather slowly with $Q^{2}$. 

In the Section 2 of this paper we review the meson cloud model, and derive 
the necessary expressions for calculating the polarised antiquark distributions. 
This will include the contributions of $\Lambda K^{*}$ and $\Sigma K^{*}$ 
components of the cloud. 
In Section 3 we present a non-perturbative calculation for the polarized parton
distributions (PDFs) of the hyperons and mesons.
The numerical results and discussions are given in Section 4.
The last section is reserved for a summary.

\section{Polarized nucleon sea in the meson cloud model}

In the meson cloud model (MCM) \cite{AThomas83} the nucleon can be viewed as
a bare nucleon plus some baryon-meson Fock states which result from
the fluctuation of nucleon to baryon plus meson $N \ra B M$.
The wavefunction of the nucleon can be written as \cite{HHoltmannSS},
\bea
|N\rangle_{\rm physical} =  Z |N\rangle_{\rm bare}
+\sum_{BM} \sum_{\lambda \lambda^\prime} 
\int dy \, d^2 {\bf k}_\perp \, \phi^{\lambda \lambda^\prime}_{BM}(y,k_\perp^2)
\, |B^\lambda(y, {\bf k}_\perp) \, M^{\lambda^\prime}(1-y,-{\bf k}_\perp)
\rangle 
\label{NMCM}
\eea
where $Z$ is the wave function renormalization constant,
$\phi^{\lambda \lambda^\prime}_{BM}(y,k_\perp^2)$ 
is the wave function of the Fock state containing a baryon ($B$)
with longitudinal momentum fraction $y$, transverse momentum ${\bf k}_\perp$,
and helicity $\lambda$,
and a meson ($M$) with momentum fraction $1-y$,
transverse momentum $-{\bf k}_\perp$, and helicity $\lambda^\prime$.
The model assumes that the lifetime of a virtual baryon-meson Fock state is much
longer than the interaction time in the deep inelastic or Drell-Yan
process, thus the quark and anti-quark in the virtual baryon-meson Fock states
can contribute to the parton distributions of the nucleon.
These non-perturbative contributions can be calculated via a convolution between
the fluctuation function, which describes the microscopic process $N \ra B M$, and
the quark (anti-quark) distributions of the hadrons in the Fock states
$|BM\rangle$ in Eq.~(\ref{NMCM}). We consider only the valence quarks of the baryon-meson
pair as the small $x$ region, where sea quarks may be important, is kinematically suppressed
(see the discussions of the fluctuation functions below). 

The contributions to the helicity-dependent parton distribution of
the nucleon are
\bea
x\delta q^\sigma &=&
~~\sum_\lambda \int_x^1 dy f^\lambda_{BM/N}(y)
	\frac{x}{y} q^\sigma_{B,\lambda}(\frac{x}{y})
+\sum_{\lambda^\prime} \int_x^1 dy f^{\lambda^\prime}_{MB/N}(y)
	\frac{x}{y} q^\sigma_{M,\lambda^\prime}(\frac{x}{y}) \nonumber \\
&&+\sum_\lambda \int_x^1 dy f^\lambda_{(B_1B_2)M/N}(y)
	\frac{x}{y} q^\sigma_{(B_1B_2),\lambda}(\frac{x}{y}) \nonumber \\
&&+\sum_{\lambda^\prime} \int_x^1 dy f^{\lambda^\prime}_{(M_1M_2)B/N}(y)
	\frac{x}{y} q^\sigma_{(M_1M_2),\lambda^\prime}(\frac{x}{y}),
\label{xq1}
\eea
\bea
x\delta \qbar^\sigma =
\sum_{\lambda^\prime} \int_x^1 dy f^{\lambda^\prime}_{MB/N}(y)
	\frac{x}{y} \qbar^\sigma_{M,\lambda^\prime}(\frac{x}{y})
+\sum_{\lambda^\prime} \int_x^1 dy f^{\lambda^\prime}_{(M_1M_2)B/N}(y)
	\frac{x}{y} \qbar^\sigma_{(M_1M_2),\lambda^\prime}(\frac{x}{y}),
\label{xqbar1}
\eea
where $f^\lambda_{BM/N}(y)$, $f^{\lambda^\prime}_{MB/N}(y)$,
$f^\lambda_{(B_1B_2)M/N}(y)$ and $f^{\lambda^\prime}_{(M_1M_2)B/N}(y)$
are the helicity dependent fluctuation functions
\bea
f^\lambda_{BM/N}(y)&=&\sum_{\lambda^\prime}\int_0^\infty dk_\perp^2
\left| \phi^{\lambda \lambda^\prime}_{BM}(y,k_\perp^2) \right|^2, \nonumber \\
f^{\lambda^\prime}_{MB/N}(y)&=&\sum_{\lambda}\int_0^\infty dk_\perp^2
\left| \phi^{\lambda \lambda^\prime}_{BM}(1-y,k_\perp^2) \right|^2, \nonumber \\
f^\lambda_{(B_1B_2)M/N}(y)&=&\sum_{\lambda^\prime}\int_0^\infty dk_\perp^2
{\phi^*}^{\lambda \lambda^\prime}_{B_1 M}(y,k_\perp^2)
\phi^{\lambda \lambda^\prime}_{B_2 M}(y,k_\perp^2), \nonumber \\
f^{\lambda^\prime}_{(M_1M_2)B/N}(y)&=&\sum_{\lambda}\int_0^\infty dk_\perp^2
{\phi^*}^{\lambda \lambda^\prime}_{B M_1}(1-y,k_\perp^2)
\phi^{\lambda \lambda^\prime}_{B M_2}(1-y,k_\perp^2).
\label{helicityff}
\eea
The last two terms in Eq.~(\ref{xq1}) and the last term in Eq.~(\ref{xqbar1})
are the interference contributions \cite{FCaoS_plzdsea}
which result from the possibility that interactions
between the photon and different baryons (mesons) can lead to the same final states.
The interference distributions $q^\sigma_{(B_1B_2),\lambda}$,
$q^\sigma_{(M_1M_2),\lambda^\prime}$ and $\qbar^\sigma_{(M_1M_2),\lambda^\prime}$ 
which do not have simple interpretations in the quark-parton model, can be related to the
PDFs of the vector mesons using SU(6) quark model wavefunctions \cite{CBorosT,FCaoS_plzdsea}.

From Eqs.~(\ref{xq1}), (\ref{xqbar1}) and (\ref{helicityff}) one can obtain
the contributions to the polarized parton distributions of the nucleon,
\bea
x\delta \Delta q &=&~~\int_x^1 d y \Delta f_{BM/N}(y)
	\frac{x}{y} \Delta q_B(\frac{x}{y})
+ \int_x^1 d y \Delta f_{VB/N}(y) \frac{x}{y} \Delta q_V(\frac{x}{y}) \nonumber\\
&&+ \int_x^1 d y \Delta f_{(B_1B_2)M/N}(y)
	\frac{x}{y} \Delta q_{(B_1B_2)}(\frac{x}{y})
+ \int_x^1 d y \Delta f_{(V_1V_2)B/N}(y)
	\frac{x}{y} \Delta q_{(V_1V_2)}(\frac{x}{y}) \nonumber \\
&&+ \int_x^1 d y f^0_{(PV)B/N}(y)
	\frac{x}{y} \Delta q_{(PV)}(\frac{x}{y}).
\label{xDeltaq}
\eea
\bea
x\delta \Delta \qbar &=&~~ \int_x^1 d y \Delta f_{VB/N}(y) \frac{x}{y}
	\Delta \qbar_V(\frac{x}{y}) \nonumber\\
&&+ \int_x^1 d y \Delta f_{(V_1V_2)B/N}(y)
	\frac{x}{y} \Delta \qbar_{(V_1V_2)}(\frac{x}{y})
+ \int_x^1 d y f^0_{(PV)B/N}(y)
	\frac{x}{y} \Delta \qbar_{(PV)}(\frac{x}{y}),
\label{xDeltaqbar}
\eea
where $V$ ($P$) indicates the meson being a vector (pseudoscalar) meson, and
\bea
\Delta f_{BV/N} &=& f^{1/2}_{BV/N}-f^{-1/2}_{BV/N} \nonumber \\
\Delta f_{(B_1 B_2)V/N} &=& f^{1/2}_{(B_1 B_2)V/N}-f^{-1/2}_{(B_1 B_2)V/N} \nonumber \\
\Delta f_{VB/N} &=& f^{1}_{VB/N}-f^{-1}_{VB/N} \nonumber \\
\Delta f_{(V_1 V_2)B/N} &=& f^1_{(V_1 V_2)B/N}-f^{-1}_{(V_1 V_2)B/N}
\eea
are the polarized fluctuation functions.
The various polarized parton distribution functions in Eqs.~(\ref{xDeltaq}) and
(\ref{xDeltaqbar}) are defined as follows:
\bea
\Delta q_{B} &=& q^\ua_{B,1/2}-q^\da_{B,1/2} = q^\da_{B,-1/2}-q^\ua_{B,-1/2}, \nonumber \\
\Delta q_{V} &=& q^\ua_{V,1}-q^\da_{V,1} = q^\da_{V,-1}-q^\ua_{V,-1}, \nonumber \\
\Delta q_{(B_1B_2)} &=& q^\ua_{(B_1B_2),1/2}-q^\da_{(B_1B_2),1/2}
			= q^\da_{(B_1B_2),-1/2}-q^\ua_{(B_1B_2),-1/2}, \nonumber \\
\Delta q_{(V_1V_2)} &=& q^\ua_{(V_1V_2),1}-q^\da_{(V_1V_2),1}
			= q^\da_{(V_1V_2),-1}-q^\ua_{(V_1V_2),-1}, \nonumber \\
\Delta \qbar_{V} &=& \qbar^\ua_{V,1}-\qbar^\da_{V,1}
			= \qbar^\da_{V,-1}-\qbar^\ua_{V,-1}, \nonumber \\
\Delta \qbar_{V_1 V_2} &=& \qbar^\ua_{(V_1 V_2),1}-\qbar^\da_{(V_1 V_2),1}
			= \qbar^\da_{(V_1 V_2),-1}-\qbar^\ua_{(V_1 V_2),-1}, \nonumber \\
\Delta \qbar_{(PV)} &=& \qbar^\ua_{(P V),0}-\qbar^\da_{(P V),0}.
\eea

For the polarized light quark sea of the proton, we consider fluctuations
$p \ra N \pi, N \rho, N \omega$ and $p \ra \Delta \pi, \Delta \rho$.
The fluctuation $p \ra \Delta \omega$ is neglected due to isospin symmetry. 
For the polarized strange sea we consider fluctuations
$p \ra \Lambda  K, \, \Sigma K $ and $p \ra \Lambda K^*, \, \Sigma K^*$.
Due to $\pi$ and $K$ mesons being pseduoscalar, the Fock states
involving these mesons do not contribute to the polarization directly.
However the interactions
between photon and different mesons could lead to the same final states,
so there are contributions from the interference effects between
$|B \pi \rangle$ and $|B \rho \, (\omega) \rangle$, and
$|B K \rangle$ and $|B K^* \rangle$
(see Eqs.~(\ref{xDeltaubar}), (\ref{xDeltadbar}) and (\ref{xDeltasbar})).
The final expressions for the polarized sea quark distributions are
\bea
x \Delta \ubar &=& \int_x^1 d y 
	\left[\frac{1}{6}\Delta f_{\rho N/N}(y)
	+\frac{2}{3}\Delta f_{\rho \Delta/N}(y)
	+\frac{1}{2}\Delta f_{\omega N/N}(y)
	+\frac{1}{2}\Delta f_{(\rho \omega)N/N}(y)\right. \nonumber \\
	&& \left. 
	+\frac{1}{6} f_{(\pi\rho)N/N}(y)
	+\frac{2}{3} f_{(\pi\rho)\Delta/N}(y)
	+\frac{1}{2} f_{(\pi\omega)N/N}(y) \right]
	\frac{x}{y}\Delta V_\rho (\frac{x}{y}),
\label{xDeltaubar} \\
x \Delta \dbar &=& \int_x^1 d y 
	\left[\frac{5}{6}\Delta f_{\rho N/N}(y)
	+\frac{1}{3}\Delta f_{\rho \Delta/N}(y)
	+\frac{1}{2}\Delta f_{\omega N/N}(y) 
	-\frac{1}{2}\Delta f_{(\rho \omega)N/N}(y)\right. \nonumber \\
	&& \left. 
	+\frac{5}{6} f_{(\pi\rho)N/N}(y) 
	+\frac{1}{3} f_{(\pi\rho)\Delta/N}(y)
	-\frac{1}{2} f_{(\pi\omega)N/N}(y) \right]
	\frac{x}{y}\Delta V_\rho (\frac{x}{y}),
\label{xDeltadbar} \\
x \Delta s &=& \int_x^1 d y 
	\left[\Delta f_{\Lambda K/N}(y) + \Delta f_{\Lambda K^*/N}(y) \right]
	\frac{x}{y} \Delta s_\Lambda(\frac{x}{y}) \nonumber \\
	&& +\left [\Delta f_{\Sigma K/N}(y) + \Delta f_{\Sigma K^*/N}(y) \right]
	\frac{x}{y} \Delta s_\Sigma(\frac{x}{y}),
\label{xDeltas} \\
x \Delta \sbar &=&\int_x^1 d y
	\left[\Delta f_{K^*\Lambda /N}(y) + \Delta f_{K^*\Sigma /N}(y)\right)
	\frac{x}{y} \Delta \sbar_{K^*} \nonumber \\
	&& + \left(f^0_{(K K^*)\Lambda /N}(y) + f^0_{(K K^*)\Sigma /N}(y)\right]
	\frac{x}{y} \Delta \sbar_{KK^*}(\frac{x}{y}).
\label{xDeltasbar}
\eea
We have employed the relations among the helicity-dependent fluctuation
functions \cite{WMelnitchoukST,FCaoS_CSB} resulting from isospin symmetry and
$SU(3)$ flavour symmetry,
\bea
\Delta f_{\rho^+n/p}&=&2\Delta f_{\rho^0 p/p}=\frac{2}{3}\Delta f_{\rho N/N},
\nonumber \\
\Delta f_{\rho^-\Delta^{++}/p}&=&\frac{3}{2}\Delta f_{\rho^0\Delta^+/p }
=3\Delta f_{\rho^+\Delta^0/p}=\frac{1}{2}\Delta f_{\rho\Delta/N},
\nonumber \\
f_{(\pi^+\rho^+)n/p}&=&2 f_{(\pi^0\rho^0)p/p}=f_{(\pi \rho)N/N},
\nonumber \\
f_{(\pi^-\rho^-)\Delta^{++}/p}&=&\frac{3}{2}f_{(\pi^0\rho^0)\Delta^+/p}
=3 f_{(\pi^+\rho^+)\Delta^0/p}=\frac{1}{2}f_{(\pi \rho)\Delta/N},
\label{ffrelations1}
\eea
\bea
\Delta f_{\Sigma^+ K^0 /p} &=& 2 \Delta f_{\Sigma^0 K^+ /p} =
	 \frac{2}{3} \Delta f_{\Sigma K /N}, \nonumber \\
\Delta f_{\Sigma^+ {K^*}^0 /p} &=& 2 \Delta f_{\Sigma^0 {K^*}^+ /p} =
	 \frac{2}{3} \Delta f_{\Sigma K^* /N}, \nonumber \\
\Delta f_{{K^*}^0 \Sigma^+ /p} &=& 2 \Delta f_{{K^*}^+ \Sigma^0/p} =
	 \frac{2}{3} \Delta f_{K^* \Sigma/N}, \nonumber \\
f^0_{(K^0 {K^*}^0)\Sigma^+ /p} &=& 2 f^0_{(K^+ {K^*}^+)\Sigma^0 /p} =
	f^0_{(K K^*)\Sigma /N}.
\label{ffrelations2}
\eea
The polarized parton distributions and `interference' distributions of different
charge states of baryons (mesons) are related using SU(6) wavefunction of
the baryons (mesons) \cite{FCaoS_plzdsea},
\bea
\Delta\dbar_{\rho^+} &=& \Delta \ubar_{\rho^-}=
2\Delta\dbar_{\rho^0}=2\Delta \ubar_{\rho^0} = \Delta V_\rho \nonumber \\
\Delta\dbar_{\rho^0\omega}&=&-\Delta\ubar_{\rho^0\omega}=
-\frac{1}{2}\Delta V_\rho,\nonumber \\
\Delta\dbar_{(\pi^+\rho^+)_{0}}&=&\Delta\ubar_{(\pi^-\rho^-)_{0}}
=2\Delta\dbar_{(\pi^0\rho^0)_{0}}=2\Delta\ubar_{(\pi^0\rho^0)_{0}}=\Delta V_\rho, 
\nonumber \\
\Delta\dbar_{(\pi^0\omega)_{0}}&=&-\Delta\ubar_{(\pi^0\omega)_{0}}=
-\frac{1}{2}\Delta V_\rho, \nonumber \\
\Delta s_{K^*} &=& \Delta s_{K^*}^+ = \Delta s_{K^*}^0, \nonumber \\
\Delta \sbar_{KK^*} &=& \Delta \sbar_{K^+ {K^*}^+} = \Delta \sbar_{K^0 {K^*}^0} 
	= \Delta s_{K^*}, \nonumber \\
\Delta s_\Sigma &=& \Delta s_{\Sigma^+} =  \Delta s_{\Sigma^0}, \nonumber\\
\Delta s_{\Lambda \Sigma^0} &=& 0.
\eea
The interference distribution $\Delta s_{\Lambda \Sigma^0}$ vanishes so there is no
interference contribution to the polarized strange sea of the nucleon (see Eq. (\ref{xDeltas})).

The wave functions $\phi^{\lambda \lambda^\prime}_{MB}(y,k_\perp^2)$
which determine the fluctuation functions (see Eq.~(\ref{helicityff}))
are calculated using time-order perturbation theory in the
infinite momentum frame (the meson being treated as on the energy-shell {\it i.e.}
$E_M=E_N-E_B$) \cite{HHoltmannSS},
\bea
\phi^{\lambda \lambda^\prime}_{BM}(y,k_\perp^2)
=\frac{1}{2\pi\sqrt{y(1-y)}}
  \frac{\sqrt{m_N m_B} \,
       V_{IMF}^{\lambda\lambda^\prime}(y,k_\perp^2) \, G_{BM}(y,k_\perp^2)}
       {m_N^2-m^2_{BM}(y,k_\perp^2)},
\label{Eq:phi_IMF}
\eea
where $m_{BM}^2$ is the invariant mass squared of the $BM$ Fock state,
\bea
m_{BM}^2(y,k_\perp^2)=\frac{m_B^2+k_\perp^2}{y}
+\frac{m_M^2+k_\perp^2}{1-y}.
\eea
The vertex function $V_{IMF}^{\lambda\lambda^\prime}(y,k_\perp^2)$ depends
on the effective interaction Lagrangian that describes the fluctuation process
$N \ra B M$. From the meson exchange model for hadron production
\cite{HHoltmannSS,RMachleidtHE} we have
\bea
{\cal L}_1 &=&i g {\bar N}\gamma_5 \pi B, \nonumber \\
{\cal L}_2 &=&  f {\bar N} \partial_\mu \pi {\Delta}^\mu + \mbox{h.c.}, \nonumber \\
{\cal L}_3 &=&  g {\bar N} \gamma_\mu\theta^\mu B
			     +f {\bar N} \sigma_{\mu\nu}
		B (\partial^\mu\theta^\nu-\partial^\nu\theta^\mu), \nonumber \\
{\cal L}_4 &=&i  f {\bar N} \gamma_5 \gamma_\mu \Delta_\nu	
	(\partial^\mu\theta^\nu-\partial^\nu\theta^\mu) + \mbox{h.c.},
\label{langragians}
\eea
where $N$ and $B=\Lambda,~\Sigma$ are spin-1/2 fields,
$\Delta$ a spin-3/2 field of Rarita-Schwinger form ($\Delta$ baryon),
$\pi$ a pseudoscalar field ($\pi$ and $K$ mesons),
and $\theta$ a vector field ($\rho$, $\omega$ and $K^*$).
The coupling constants for various considered fluctuations
are taken to be \cite{BHolzenkampHS,RMachleidtHE},
\bea
g^2_{NN\pi}/4\pi = 13.6,& &  \nonumber \\
g^2_{NN\rho}/4\pi=0.84, & & f_{NN\rho}/g_{NN\rho}=6.1/4m_N, \nonumber \\
g^2_{NN\omega}/4\pi=8.1, & & f_{NN\omega}/g_{NN\omega}=0, \nonumber \\
f^2_{N\Delta\pi}/4\pi=12.3~{\rm GeV}^{-2}, & &
f^2_{N\Delta\rho}/4\pi=34.5~{\rm GeV}^{-2}, \nonumber \\
g_{N\Lambda K} =  -13.98, & &
g_{N\Sigma K}  =   2.69,  \nonumber \\
g_{N\Lambda K^{*}}  = -5.63, & & 
f_{N\Lambda K^{*}}  = -4.89~{\rm GeV}^{-1}, \nonumber \\
g_{N\Sigma K^{*}}  = -3.25, & & 
f_{N\Sigma K^{*}}  = 2.09~{\rm GeV}^{-1}.
\label{couplingconstant}
\eea

The phenomenological vertex form factor $G_{BM}(y,k_\perp^2)$ in Eq.~(\ref{Eq:phi_IMF})
is introduced to describe the unknown dynamics of the fluctuation $N\ra BM$,
for which we adopt the exponential form
\bea
G_{BM}(y,k_\perp^2)={\rm exp} \left[\frac{m_N^2-m_{BM}(y,k_\perp^2)}
{2\Lambda_C^2}\right],
\label{Eq:GBM}
\eea
with $\Lambda_C$ being a cut-off parameter.
This form factor satisfies the relation
$G_{BM}(y,k_\perp^2)=G_{MB}(1-y,k_\perp^2)$.

\section{Polarized parton distribution functions of hadrons}

The polarized parton distribution functions of the hyperons $\Lambda$ and $\Sigma$ and
mesons $\rho$, $\omega$ and $K^*$ are largely unknown.
In order to estimate these distributions we extend the method of 
the Adelaide group \cite{Bag_Adelaide} which uses the bag model
to evaluate the parton distributions of baryons.
The bag model calculations give results consistent with the experimental data for
the parton distributions of the nucleon, and
the calculations have been extended to other baryons \cite{CBorosT}.
As the bag model gives a good description of many non-perturbative properties
(e.g. the mass spectrum) of the mesons except for the pion \cite{DeGrand75},
we argue that bag model calculations of the parton distributions
of the mesons should give a reasonably good approximation to these distributions.

In the bag model the dominant contributions to the parton distribution functions of
a hadron in the medium-$x$ range come from intermediate states with the lowest
number of quarks, so the intermediate states we consider contain one quark
(or anti-quark) for the mesons and two quarks for the hyperons.
Following \cite{Bag_Adelaide} we can write these contributions as
\bea
q_{h,f}^{\ua \da}(x) = 
\frac{M_{h}}{(2\pi)^{3}} \sum_{m} \langle \mu | P_{f,m}| \mu \rangle 
\int d{\bf p}_{n} \frac{|\phi_{i}({\bf p}_{n})|^{2}}{|\phi_{j}({\bf 0})|^{2}} 
\delta(M_{h}(1-x) - p_{n}^{+}) |\tilde{\Psi}^{\ua \da}_{+,f}({\bf p}_{n})|^{2},
\label{mit2q}
\eea
where $M_h$ is the hadron mass,
`+' components of momenta are defined by $p^+ = p^0 + p^3$, and
${\bf p}_{n}$ is the 3-momentum of the intermediate state.
$\tilde{\Psi}$ is the Fourier transform of the MIT bag ground state wavefunction
$\Psi({\bf r})$, and $\phi_{m}({\bf p})$ is the Fourier transform of the Hill-Wheeler 
overlap function between $m$-quark bag states:
\bea
|\phi_{m}({\bf p})|^{2} = \int d{\bf R} e^{-i{\bf p \cdot R}}
\left[ \int d{\bf r} \Psi^{\dagger}({\bf r-R}) \Psi({\bf r}) \right]^{m}.
\eea
In Eq.~(\ref{mit2q}) one takes $i=1, \, j=2$ for the mesons
($\rho, \, \omega,\, K,$ and $K^*$)
and $i=2, \, j=3$ for the hyperons ($\Lambda$ and $\Sigma$).
The matrix element $\langle \mu | P_{f,m}| \mu \rangle$ appearing in  
Eq.~(\ref{mit2q}) is the matrix element of the projection operator 
$P_{f,m}$ onto the required flavour $f$ and helicity $m$ for the $SU(6)$ 
spin-flavour wavefunction $| \mu \rangle$ of the hadron under consideration.

The input parameters in the bag model calculations are 
the bag radius $R$, the mass of the quark (anti-quark) $m_{q}$
for which the parton distribution is calculated, the mass of the intermediate state $m_{n}$,
and the bag scale $\mu^{2}$ - at this scale the model is taken 
as a good approximation to their valence structure of the hadron. In Table~I
we list the values for these parameters adopted in this work.
In figure 1 we show the polarized parton distribution functions after
NLO evolution \cite{Evolution} to the scale of $Q^2=2.5$ GeV$^2$
where the HERMES results are available.
As expected, the PDFs of the mesons are harder than those of the hyperons since
the dominant intermediate states of the mesons containing one quark (anti-quark)
are lighter than those of the hyperons containing two quarks.
Also we can see that $x\Delta s_{K^*}$ is harder than $x \Delta V_\rho$
due to the $s$-quark being heavier than the $u$- and $d$-quarks.
The polarization of $s$-quark in the $\Lambda$ hyperon is positive while it is negative for
the $\Sigma$ hyperon since the SU(6) wavefunction of the $\Sigma$ is dominated
by the term $ | u^\uparrow d^\uparrow s^\downarrow \rangle $.
The polarized PDF of strange quarks in the $\Lambda$ ($x \Delta s_\Lambda$) is harder 
than that of the $\Sigma$ ($x \Delta s_\Sigma$) because the two-quark intermediate state
for the $\Lambda$ is a light scalar while it is a vector for the $\Sigma$ which is
heavier by $200$ MeV because of the hyperfine splitting between $qq$ states.

\section{Numerical Results and Discussions}

The fluctuation functions depend on the cut-off parameter $\Lambda_C$ introduced in the
phenomenological vertex form factor $G_{BM}$ (see Eq.~(\ref{Eq:GBM})).
We adopt $\Lambda_C^{\rm oct}=1.08$ GeV and $\Lambda_C^{\rm dec}=0.98$ GeV
for the fluctuations involving the octet and decuplet baryons respectively 
\cite{HHoltmannSS}. These values are determined from analysis of high energy
$p$-$p$ and $p$-$\Lambda$ scattering, and give a reasonable fit to $\dbar(x) - \ubar(x)$
in the unpolarized nucleon sea \cite{HHoltmannSS}.
The polarized fluctuation functions needed for calculating the polarized strange sea of
the nucleon (see Eqs.~(\ref{xDeltas}) and (\ref{xDeltasbar})) are shown in figure 2.
For the fluctuation functions involved in calculating the polarized
light quark sea of the nucleon ($\Delta \ubar, \, \Delta \dbar$)
we refer to \cite{FCaoS_plzdsea}.
We note that the fluctuation functions $\Delta f_{\Lambda K^*/N}$ and $\Delta f_{\Sigma K^*/N}$
are larger in magnitude than $\Delta f_{\Lambda K/N}$ and $\Delta f_{\Sigma K/N}$, although
one might expect that the $K^*$ fluctuation functions would be smaller than the
$K$ fluctuation functions due to the higher mass of the $K^*$.
Also the $K^*$ fluctuation functions are negative while the $K$ fluctuation functions
are positive, 
so the calculation of $\Delta s$ is sensitive to whether contributions from
fluctuations involving $K^*$ are included or not.
In the case of $\Delta \sbar$, the contributions from $K^*$ states
are the leading contributions in the MCM.
The sum of the fluctuation functions $f^0_{(KK^*)\Lambda/N}$ and $f^0_{(KK^*)\Sigma/N}$
changes sign and is much smaller in magnitude than the sum of $\Delta f_{K^*\Lambda/N}$ and 
$\Delta f_{K^*\Sigma/N}$, which indicates that the contribution from the $K$-$K^*$ interference
is not significant. The same conclusion is also true in the calculation for
the light polarized quark sea \cite{FCaoS_plzdsea}.
 
The results for the light quark sea $x\Delta \ubar(x)$ and $x\Delta \dbar(x)$,
along with the data from the HERMES collaboration, are presented in figure 3.
The calculations show that the polarizations in the light quark sea are positive
and the polarization of the anti-up quark is about $10\%$ of
that of the anti-down quark.
Thus the SU(2) flavor symmetry ($\Delta \ubar(x) = \Delta \dbar(x)$)
in polarized light quark sea is broken and $\Delta \ubar(x) < \Delta \dbar(x)$
over the range of $0.01 < x < 0.6$.
The calculations for $x\Delta \ubar(x)$ and $x\Delta \dbar(x)$ are consistent with the data.
To highlight the flavour symmetry breaking, we calculate the difference
$x(\Delta \ubar(x) -\Delta \dbar(x))$ and compare it with the HERMES result in figure 4.
Our theoretical calculations are consistent with the data, although large uncertainties exist
in the data.
Also we can find the SU(2) flavour symmetry breaking in the polarized nucleon sea is much smaller
than in the unpolarized sea, which is in contrast to calculations using chiral quark soliton
model \cite{Chiral} which predict the differences $(\Delta \dbar - \Delta \ubar)$
and $(\dbar - \ubar)$ are similar in magnitude.

The contributions to the light polarized antiquark distributions calculated in this work
come mainly from the antiquark in the meson cloud.
There may be other non-perturbative contributions to flavour symmetry breaking of
the parton distribution of the bare nucleon.
Some studies \cite{FCaoS_plzdsea,PauliBlocking}
estimated that these contributions could be significantly larger than the contributions
from the meson cloud by considering Pauli blocking effects.
However the HERMES data indicate that these
non-perturbative contributions from the bare nucleon cannot be very large.
As Pauli blocking effects are expected to be of similar size in both polarized and unpolarized
case \cite{FCaoS_plzdsea,PauliBlocking}, this conclusion may be of important in discussions of
$\dbar - \ubar$ difference \cite{Explanations}

In figure 5 we show the polarization of the strange sea calculated both with and without
the contributions from Fock states involving $K^*$ mesons.
We can see that the predictions depend strongly on contributions from the $K^*$ Fock states.
We have arrived at a similar conclusion on the importance of considering the $K^*$ mesons
in a recent investigation of the unpolarized strange sea \cite{FCaoS_KKstar}.
To study the quark-antiquark symmetry breaking in the polarized strange sea
we show the difference $x(\Delta s(x) - \Delta \sbar(x))$ in figure 6.
It can be seen that $x(\Delta s(x) - \Delta \sbar(x))< 0$ 
when both contributions from $K$ and $K^*$ mesons are included, while
$x(\Delta s(x) - \Delta \sbar(x))>0$ when only $K$ mesons are considered.
In figure 7 we compare theoretical calculations for $x(\Delta s(x) + \Delta \sbar(x))/2$
with the HERMES measurement for $x \Delta s(x)$.
It is interesting to compare the strange-antistrange asymmetry in the polarized sea with
that in the unpolarized sea. We present such a comparison in figure 8. We find that
the strange-antistrange asymmetry is much more significant in the polarized sea 
than in the unpolarized sea.

The integrals of polarized parton distribution functions ($\Delta Q=\int^1_0 \Delta q(x) dx$)
give the contribution to the spin of the nucleon carried by each flavor of parton.
We found that $\Delta \bar{U}=0.001$ $\Delta \bar{D}=0.03$
and $\Delta S +\Delta \bar{S}=0.01 \, (0.004)$ with (without) the $K^*$ Fock states.
The total spin carried by charged partons ($\Delta \Sigma$) is determined by DIS experiments
to be about $0.3$, so the light antiquark sea and strange quark sea contribute about $10\%$ of
this total spin.
The polarization of the strange quark sea is found to be positive which is in contradiction to
the previous conclusion that the strange quark and anti-quarks are polarized negatively
with respect to the direction of the nucleon spin ($\Delta S + \Delta \bar{S} < 0$)
based on analyses of inclusive deep inelastic scattering (DIS) \cite{IDIS}
and lattice calculations \cite{Lattice}.
Also our result for $\Delta S + \Delta \bar{S}$ is about $10\%$ of the magnitude found in
previous analyses.
However our prediction of a positively polarised strange sea $(\Delta S + \Delta \bar{S})$
agrees with the HERMES result,
$\int_{0.023}^{0.3} \Delta s(x) dx=0.03 \pm 0.03 (\rm stat.) \pm 0.01 (\rm sysrt.)$.

\section{Summary}
The polarized parton distribution functions of the nucleon sea provide vital information on the
non-perturbative structure of the nucleon.
In this paper we have calculated the polarized parton distribution functions of the nucleon sea 
using the meson cloud model and thereby investigated the flavour and quark-antiquark symmetries
of the nucleon. Our calculations show that the SU(2) flavour symmetry
and quark-antiquark symmetry in the polarized nucleon sea are broken
and $\Delta \ubar(x) < \Delta \dbar(x)$ and $\Delta s(x) < \Delta \sbar(x)$.
SU(2) flavour symmetry breaking in the polarized nucleon sea is found to be
much smaller than in the unpolarized sea. This is in contrast to calculations
in the chiral quark soliton model, or calculations based on
Pauli blocking, which have found $(\Delta \dbar - \Delta \ubar)$
to be similar in magnitude to $(\dbar - \ubar$).

The strange-antistrange symmetry breaking is much larger in the polarized nucleon than
in the unpolarized nucleon.
Our finding of a slightly positively polarized strange sea is remarkably different
from previous determinations of a significantly negatively polarized strange sea. 
This may be due to a breaking of SU(3) flavour symmetry, e.g. the F and D values
calculated from hyperon decays may not apply to the nucleon.
The contributions to the total spin carried by the charged partons from the light antiquark sea
and strange sea is about $10\%$.
Our calculations generally agree with recent results from the HERMES Collaboration
though large error bars exist in the data.
More experimental data with high precision are highly desired and will put more rigorous constraints
on models of nucleon structure.

\section*{Acknowledgments}
We thank M. Beckmann for providing numerical results of HERMES measurement
and useful discussions on the data.
This work was partially supported by the Science and Technology Postdoctoral
Fellowship of the Foundation for Research Science and Technology, and the 
Marsden Fund of the Royal Society of New Zealand.

\newpage
\section*{Figure Captions}
\begin{description}
\item
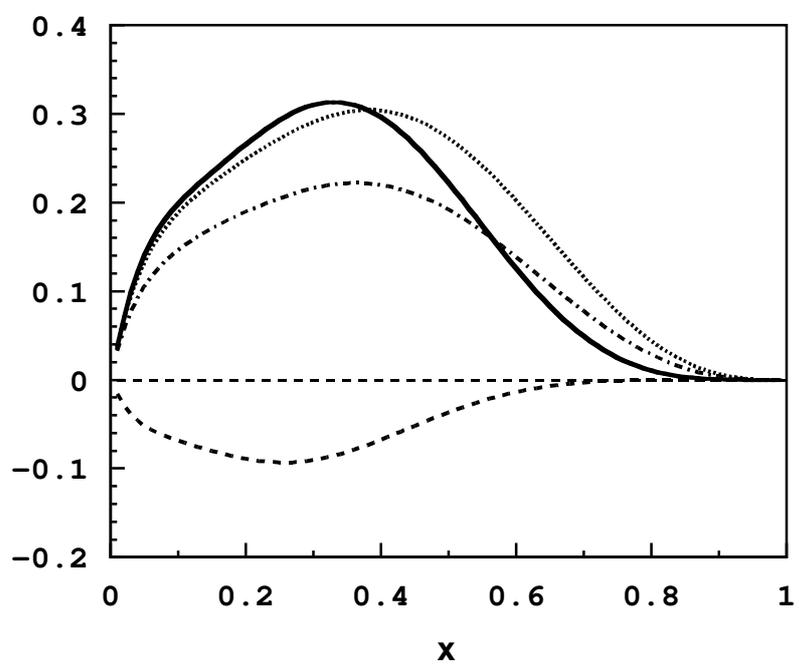
{Fig.~1.}
Bag model calculations for the polarized parton distribution functions
of hyperons $\Lambda$ ($x \Delta s_\Lambda$, solid curve) and
$\Sigma$ ($x \Delta s_\Sigma$, dashed curve)
and mesons $K^*$ ($x \Delta \sbar_{K^*}$, dotted curve) and
$\rho$ ($x \Delta V_\rho$, dash-dotted curve).
All distributions are evolved to the scale of $Q^2=2.5$~GeV$^2$.
\item
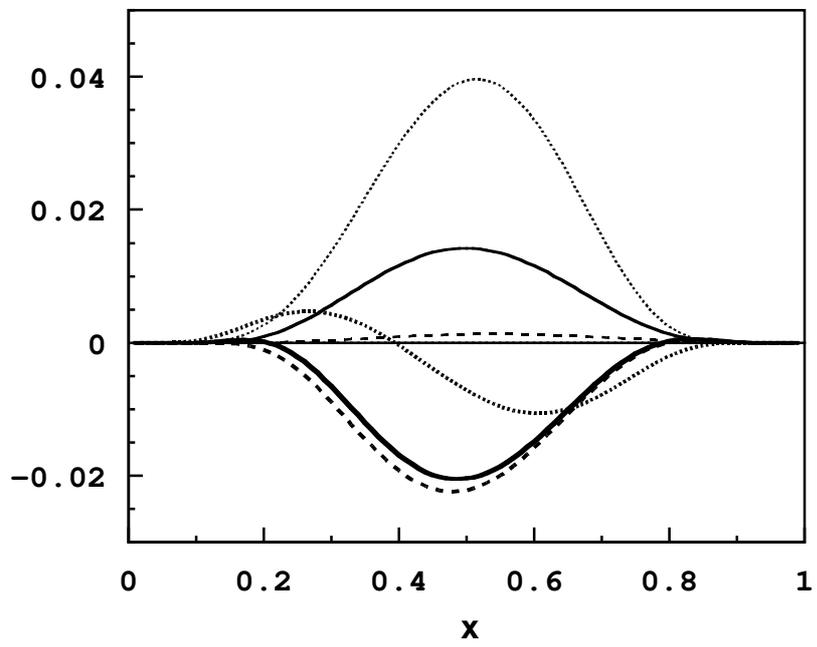
{Fig.~2.}
The polarized fluctuation functions. The thin solid and dashed curves are for
$\Delta f_{\Lambda K/N}$ and $\Delta f_{\Sigma K/N}$ respectively
while the thick solid and dashed curves are for $\Delta f_{\Lambda K^*/N}$
and $\Delta f_{\Sigma K^*/N}$ respectibvely.
The thin and thick dotted curves are for $\Delta f_{K^*\Lambda/N}+\Delta f_{K^*\Sigma/N}$
and the interference term $f^0_{(KK^*)\Lambda/N} + f^0_{(KK^*)\Sigma/N}$.
\item
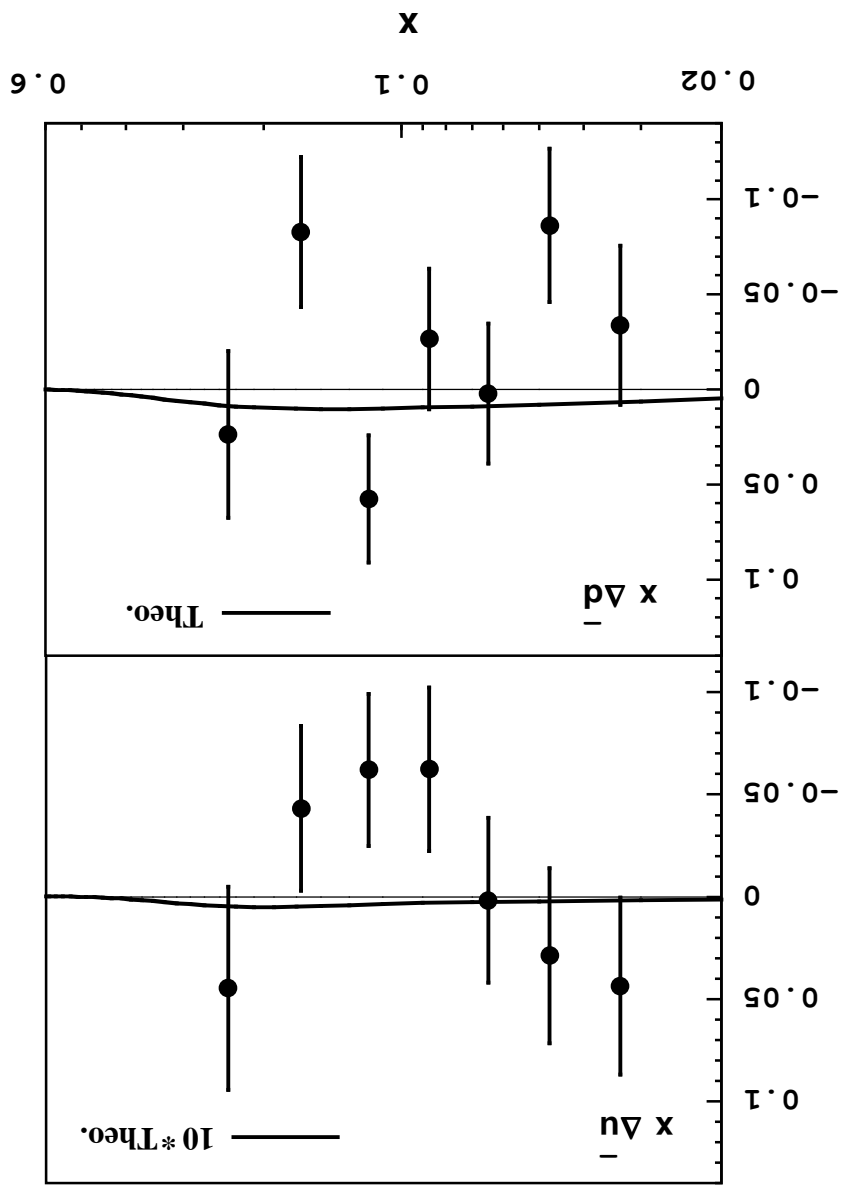
{Fig.~3.}
Comparison of theoretical calculations for the polarized light anti-quark sea and
the experimental data from the HERMES collaboration \cite{HERMES02}.
\item
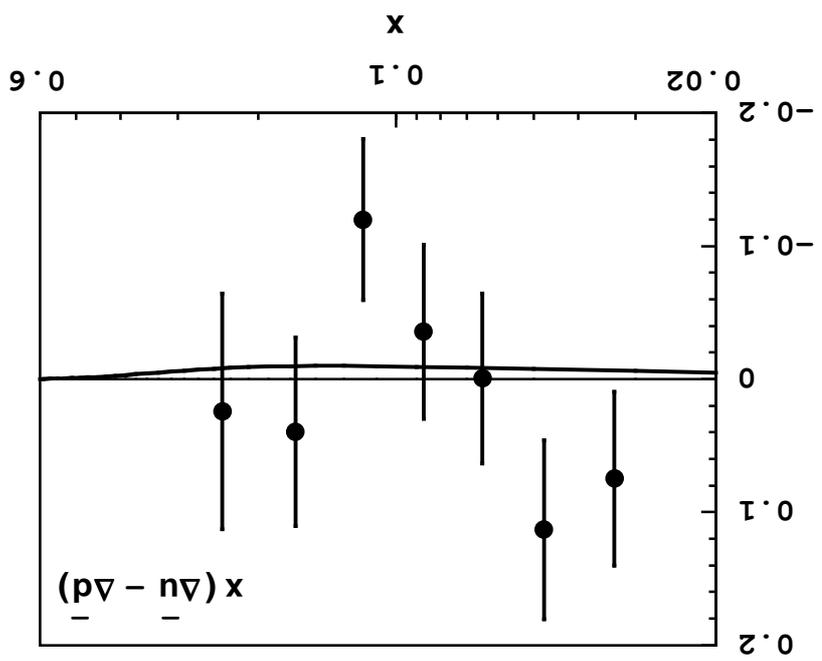
{Fig.~4.}
Flavour asymmetry in the polarized light anti-quark sea.
The solid curve is the theoretical calculation.
The HERMES data are taken from \cite{HERMES02}.
\item
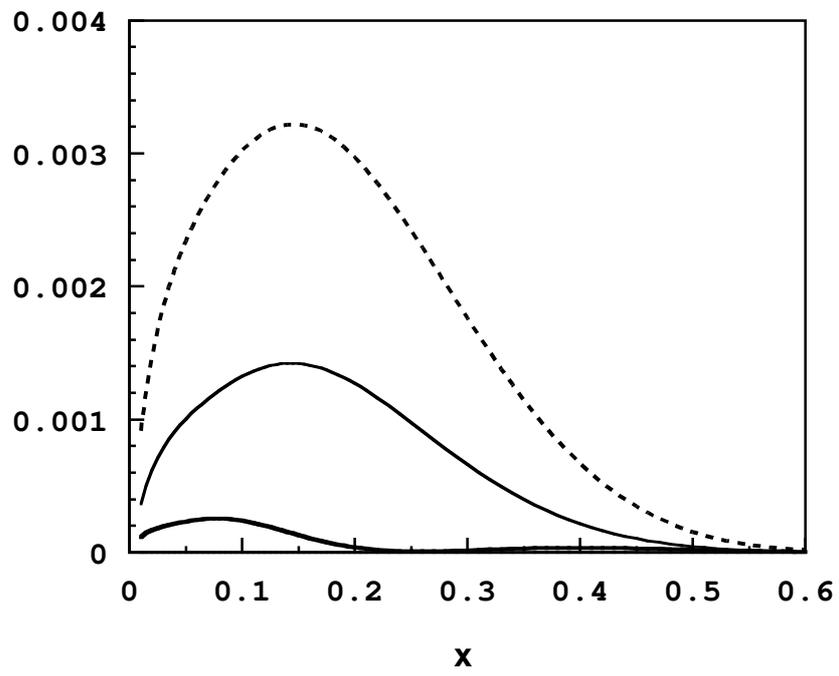
{Fig.~5.}
Polarized strange sea of the nucleon.
The thin solid curve is theoretical calculation for $x \Delta s$ when only $K$ Fock states
being considered.
The thick solid and dashed curves are results for $x \Delta s$ and $x \Delta \sbar$
including the contributions from both $K$ and $K^*$ Fock states.
\item
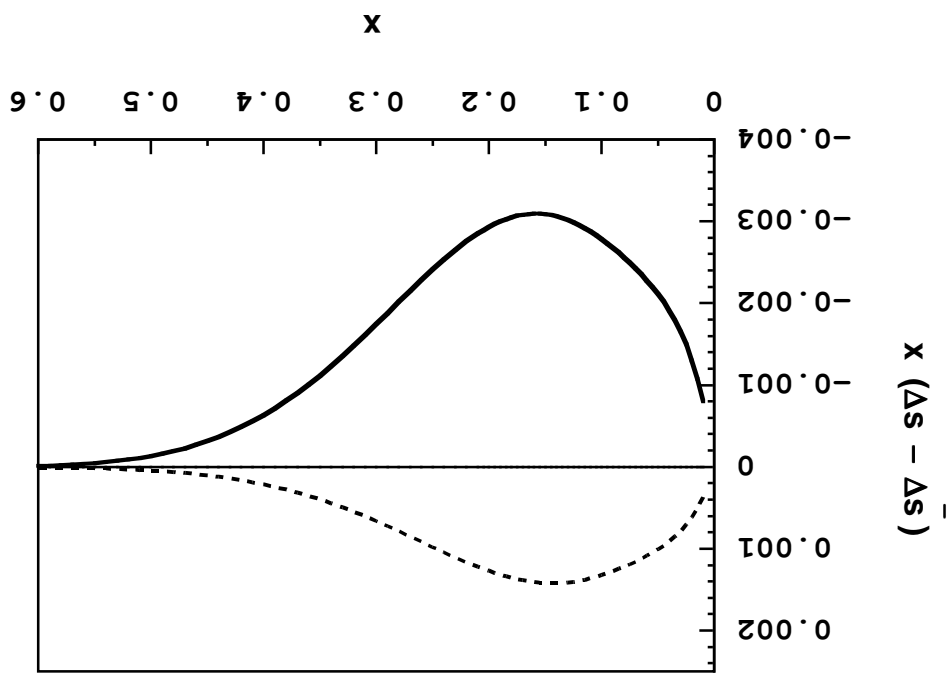
{Fig.~6.}
Strange-antistrange asymmetry $x(\Delta s- \Delta \sbar)$ in the polarized nucleon sea.
The solid curve is the result including the contributions from both $K$ and $K^*$ Fock states,
while the dashed curve is the result including only the $K$ Fock states.
\item
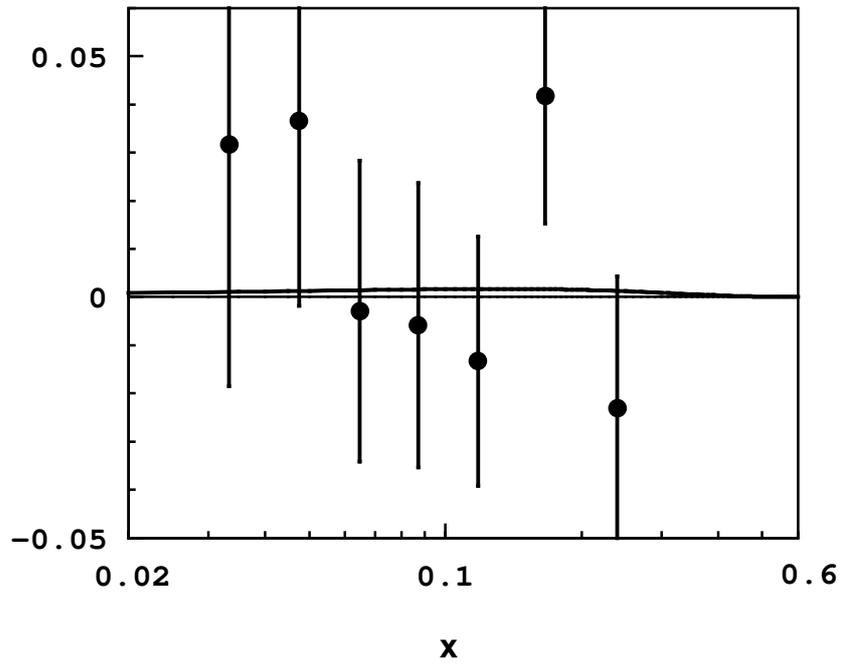
{Fig.~7.}
Comparison of theoretical calculations for $x(\Delta s + \Delta \sbar)/2$ with
the HERMES results for $x \Delta s(x)$ \cite{HERMES02} at $Q^2=2.5$~GeV$^2$.
\item
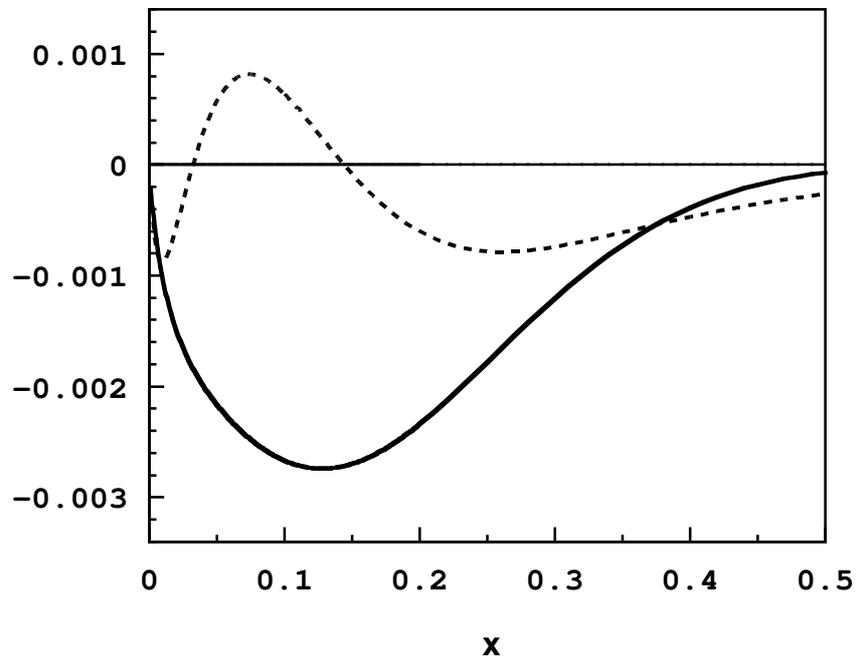
{Fig.~8.}
Strange-antistrange asymmetry in the unpolarized and polarized sea.
$x( s- \sbar)$ at $Q^2=16$~GeV$^2$, dashed curve,
and $x(\Delta s- \Delta \sbar)$ at $Q^2=2.5$~GeV$^2$, solid curve.

\end{description}

\newpage
\begin{center}
Table I. Input parameters in the bag model calculation.
\vskip 0.5cm
\begin{tabular}{|c|c|c|c|c|}\hline
          & $R$(fm) & $m_q$(MeV) & $m_n$(MeV) & $\mu^2$(GeV$^2$) \\ \hline     
$\Lambda$ & $0.8$   & $150$      & $650$      & $0.23$ \\ \hline
$\Sigma$  & $0.8$   & $150$      & $850$      & $0.23$ \\ \hline
$\rho$    & $0.7$   & $0$        & $425$      & $0.23$ \\ \hline
$K^\star$ & $0.7$   & $150$      & $425$      & $0.23$ \\ \hline
\end{tabular}
\end{center}

%

\end{document}